\documentclass[fleqn]{elsart}
\usepackage{graphicx}
\usepackage{amsmath}
\usepackage{amsfonts}
\usepackage{amssymb}

\begin{document}

\begin{frontmatter}
\title{Monte Carlo simulation study of exchange biased hysteresis loops in nanoparticles}

\author{\`{O}scar Iglesias}
\ead{oscar@ffn.ub.es}
\ead[url]{http://www.ffn.ub.es/oscar}
and 
\author{Am\'{\i}lcar Labarta} 
\address{Departament de F\'{\i}sica Fonamental, Universitat de Barcelona, Diagonal 647, 08028 Barcelona, Spain}

\begin{abstract}
We present the results of Monte Carlo simulations of the magnetic properties of a model for a single nanoparticle consisting in a ferromagnetic core surrounded by an antiferromagnetic shell. The simulations of hysteresis loops after cooling in a magnetic field display exchange bias effects. In order to understand the origin of the loop shifts, we have studied the thermal dependence of the shell and interface magnetizations under field cooling. These results, together with inspection of the snapshots of the configurations attained at low temperature, show the existence of a net magnetization at the interface which is responsible for the bias of the hysteresis loops.
\vspace{1pc}
\end{abstract}

\end{frontmatter}  
\section{Introduction}
In spite of its first report in the 50's \cite{Meiklejohn_PR}, exchange bias effects in nanoparticle systems are still an active subject of research nowadays because of its technological application \cite{Reviews}. Current synthesis techniques have allowed the preparation of granular and particulate magnetic systems, in which ferromagnetic particles have either oxide shells or are coupled to AF environments. Shifts of the hysteresis loops after field cooling, similar to those observed in layered strucutures, have been experimentally observed in these systems \cite{NoguesN}. However, despite the similarities in both cases, the theoretical models used to account for this effect in layered systems \cite{Kiwi} are not well suited for particle systems. Surface effects due to the reduced dimensionality of the nanoparticles and the exchange coupling at the FM-AFM interfaces are supposed to play a role in the observation of exchange bias that can be properly studied by Monte Carlo simulations of a single particle \cite{Iglesias_PhysicaB}.

\section{Model}

We have performed Monte Carlo simulations of a model of a nanoparticle consisting of Heisenberg spins $\vec{S}_i$ on a simple cubic lattice. The particle shape is spherical with radius $R= 12 a$ (in units of the cell size $a$). The particle outer shell of spins having a width $R_{Sh}= 3 a$ is considered as the surface and we will assign its spins magnetic properties different from those at the core. In particular, we are interested in a particle with a ferromagnetic (FM) core with uniaxial anisotropy $k_C$ along the long axis surrounded by a shell of pins with antiferromagnetic (AF) interactions and also uniaxial anisotropy $k_S$.  
The interaction Hamiltonian considered can be written as:
\begin{eqnarray}
\label{Eq1}
{ H}/k_{B}= 
-\sum_{\langle  i,j\rangle}J_{ij}{\vec S}_i \cdot {\vec S}_j   
-k_C\sum_{i\in \mathrm{C}}(S_i^z)^2\nonumber\\
-k_S\sum_{i\in \mathrm{Sh}}(S_i^z)^2  	
-\sum_{i= 1}^{N} \vec h\cdot{\vec S_i}
\end{eqnarray}
where $\vec{h}$ is the magnetic field ($\vec{h}=\mu\vec{H}/k_B$ will denote the field strength in temperature units, with $\mu$ the magnetic moment of the spin). The first term accounts for the nearest-neighbor exchange interactions, with different exchange constants. Core spin interactions are FM ($J_{\mathrm{C}}>0$) whereas spins in the shell sorrounding the core have AF interactions ($J_{\mathrm{S}}<0$) with a reduced value with respect to that in the core in order to reproduce a Ne\'el temperature lower than the Curie temperature of the core as in real core/shell oxidized nanoparticles \cite{Peng_PRB}. Finally, in order to study the role of  FM and AF interactions across the core/shell interface, for spins at the interface [i.e., the core (shell) spins having nearest neighbors in the shell (core)] the exchange constant is $J_{\mathrm{Int}}\lessgtr 0$. 

\section{Results}

As a firt step towards the understanding of the exchange bias effects in magnetic particles, we have started by simulating the field cooling process previous to the hysteresis loop. For this purpose, we start from a disordered configuration at a temperature higher than the Ne\'el temperature of the antiferromagnetic shell and progressively reduce it in constant steps $\delta T = 0.1$ down to the measuring temperature of the hysteresis loop in the presence of a reduced magnetic field $h= 4$ K pointing along the positive z axis. At each temperature, the system is allowed to equilibrate during a number of Monte Carlo (MC) steps and after the magnetization is averaged over at least $10000$ MC steps using usual heat bath dynamics for continuous spins. The thermal dependence of the projection of the magnetization along the field direction is shown in Fig. 1, where the contributions of the spins in the core, in the shell and at the interface to the total magnetization have been plotted separately. 

We note that on field cooling the core magnetization tends to $1$ indicating a progressive alignement of all the core spins towards the field direction driven by the FM coupling between them. In a similar way, the spins at the shell attain AFM order although with a small net magnetization originating from the uncompensated spin at the shell surfaces as can be more clearly seen in the snapshot presented in Fig. 2a. More interesting is the thermal dependence of the interface magnetization, which reaches a value of $0.37$. This is an indication of the existence of uncompensated spins at the interface between the FM core and the AF shell of the particle. As it can be seen in the b-d panels of Fig. 2, the geometric structure and magnetic ordering of the interface in a core/shell nanoparticle is more intricate than in the case of FM/AF coupled bilayers with a roughness inherent to the geometry of the interface. For a nanoparticle, the core (shell) spins at the interface have different number of neighbors (1, 2 or 3 shown in panels b, c, d of Fig. 2 respectively) on the shell (core) depending on their position. Moreover, the interface spins at the shell present regions with either local compensated or uncompensated magnetic order.

Considering the configurations obtained after the previously described field cooling process as a starting point, we have also simulated hysteresis loops by cycling the magnetic field between the $\pm h_{\mathrm{FC}}$ values at $T= 0.1$. Along the loops, the field is varied in constant steps $\delta h= 0.2$ during which averages of the magnetization are performed during $100$ MCS after other initial $100$ MCS are discarded for thermalization. The parameters that determine the characteristics of the hysteresis loops are the anisotropy constants at the core (which as been set to $K_\mathrm C=1$) and at the shell $K_\mathrm{Sh}$, and the exchange constants at the core ($J_\mathrm C= 10$, which fixes the Curie temperature $T_C$ of the FM), at the shell $J_\mathrm{Sh}$ and at interface $J_\mathrm{Int}$. 
It turns out that disorder and frustration at the surface induced by radial anisotropy and finite-size effects alone are not enough to produce sizable loops shifts as simulations performed for particles with no AF shell demonstrate \cite{Iglesias_PRB01}. Therefore, as most experiments in which exchange bias has been observed, the particle shell is in an oxidized state with Ne\'el temperature below $T_C$, so that we have set $J_\mathrm{Sh}= -0.5 J_\mathrm{C}$. Moreover, enhancement of the anisotropy at the surface as compared to that at the core is necessary to pin the shell spins into the cooling field direction so that a shift of the loops occurs after field cooling, so that we have set $K_\mathrm{Sh}= 10$), which also in agreement with the reported enhanced surface anisotropies due reduced local coordination at the outer particle shells \cite{Tronc_jmmm03,Puerto}. 

As an illustrative case, we show in Fig. 3 the hystersis loops obtained by starting from a demagnetized state obtained after cooling the particle in zero field (ZFC) and after cooling in a magnetic field $h_{\mathrm{FC}}= 4$ (FC) for FM and AF coupling of the interface spins. As it is evident from the figures, the FC loops are shifted towards negative field values with respect to the ZFC loops and show a slightly increased coercivity independently of the sign of $J_{\mathrm {Int}}$. Changing the sign of the interface coupling affects the kind of magnetic order at the interface due to the direct coupling of the core to the shell through the exchange interaction ($J_{\mathrm{Int}}$), as is reflected by the different sign of the net interface magnetization just after the FC process (see lower panels in Fig. 3). 
For FM interface coupling ($J_{\mathrm{Int}}= + 0.5 J_{\mathrm{Sh}}$), the net interface magnetization after FC partially follows the reversal of the core.
In contrast, for AF interface coupling ($J_{\mathrm{Int}}= - 0.5 J_{\mathrm{Sh}}$), $M_{\mathrm{Int}}$ remains always negative along the hysteresis loop as the magnetic field is not able to produce a positive net magnetization due to the strong AF coupling of the shell spins to the core. However, in both cases, the shift of the interface loops in the vertical direction indicates the presence of a fraction of spins that remain pinned during the reversal process and that are responsible for the observed shift of the hysteresis loops of the whole particle.

\section*{Acknowledgements}
We acknowledge CESCA and CEPBA under coordination of C$^4$ for the computer facilities. This work has been supported by the Spanish spanish MEC through the MAT2003-01124 project and the Generalitat de Catalunya through the 2001SGR00066 DURSI project.


\newpage
\section*{FIGURE CAPTIONS}
Fig. 1: (Color online) Thermal dependence of the magnetization of a core/shell particle when cooling form a disordered state at $T>T_N$ down to $T = 0.1$ in the presence of an external magnetic field $h_{\mathrm FC}= 4$ K. The different curves correspond to the contributions of the core, shell and interface spins to the total magnetization $M_{\mathrm{Total}}$.

Fig. 2: (Color online) (a) Spin configuration of an equatorial cut of the particle attained after the field cooling process described in Fig. 1. Core spins are dark blue, spins at the shell are green while inteface core and shell spins have been colored in light blue and yellow. (b-d) Configurations of the core (shell) spins at the interface having 1, 2 or 3 nearest neighbors in the shell (core).

Fig. 3: (Color online) Main panels: hysteresis loops for a particle with $R_{Sh}= 3$a and radius $R= 12$a obtained from a demagnetized state (ZFC) and after cooling in a magnetic field $h_{FC}= 4$. Lower panels: contribution of the interface spins of the shell to the total magnetization.  
The parameters used are: $J_{\mathrm C}= 10$, $J_{\mathrm S}= -0.5 J_{\mathrm C}$, $K_{\mathrm C}= 1$, $K_{\mathrm S}= 10$ and $J_{\mathrm {Int}}= -0.5\ (+0.5)$ in the left (right) panels. 
\newpage
\
\begin{figure}[tbp] 
\centering 
\includegraphics[width=\textwidth]{FC_Js-05_Ji-05.eps}
\label{FC_fig}
\end{figure}
\newpage
\
\begin{figure*}[tbp] 
\centering 
\includegraphics[width=\textwidth,angle= -90]{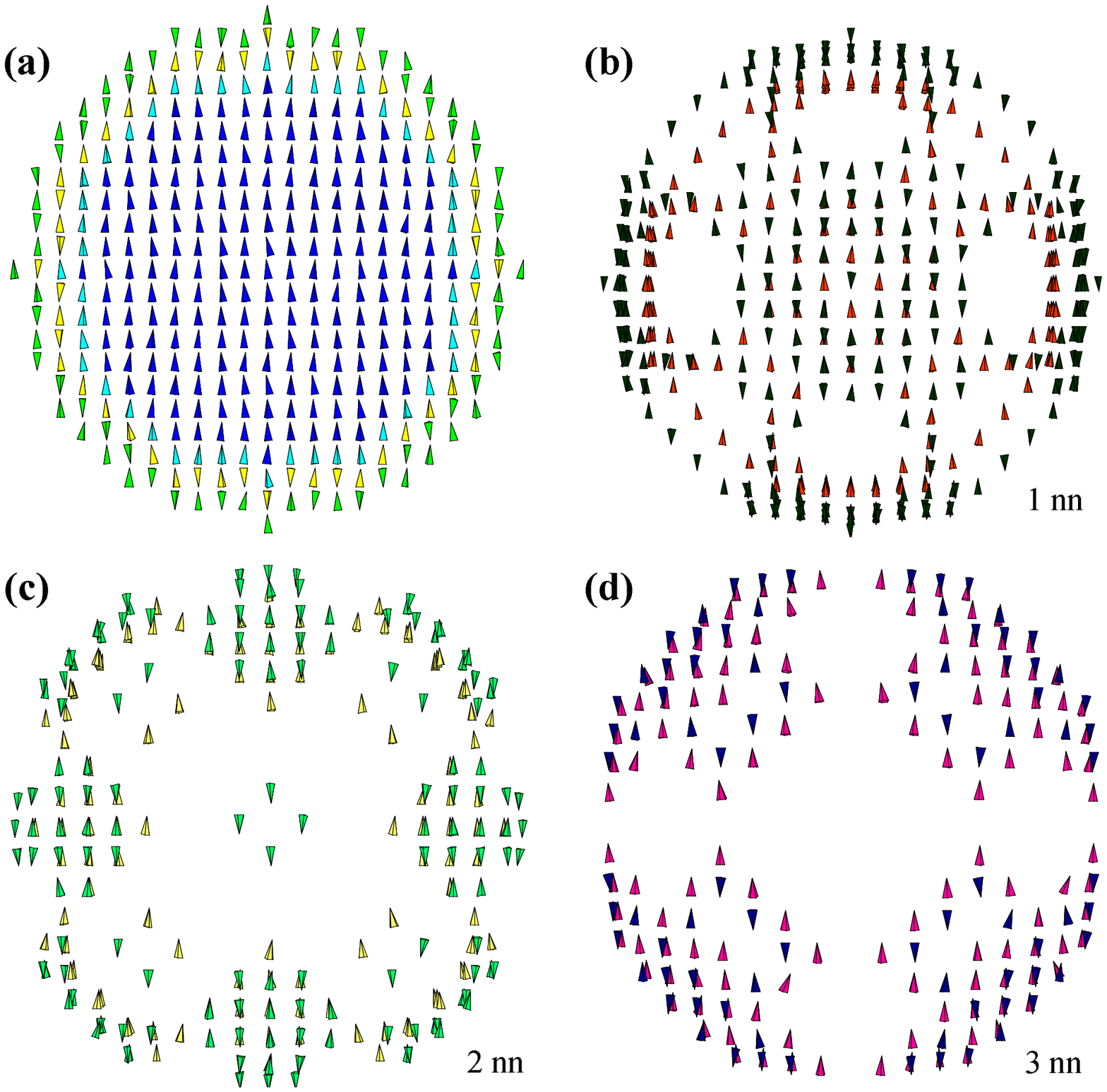}
\label{FC_Confs_fig}
\end{figure*}
\newpage
\
\begin{figure*}[tbp] 
\centering 
\includegraphics[width=0.49\textwidth]{FC+ZFC_Js-05_Ji-05_Int.eps}
\includegraphics[width=0.49\textwidth]{FC+ZFC_Js-05_Ji+05_Int.eps}
\label{HLoops_Fig}
\end{figure*}

\end{document}